%% file: main.tex
\documentclass[lettersize,journal]{IEEEtran}

\usepackage{cite}
\usepackage{amsmath,amssymb,amsfonts}
\usepackage[ruled,vlined]{algorithm2e}
\usepackage{graphicx}
\usepackage{textcomp}
\usepackage{url}
\usepackage{booktabs}
\usepackage{multirow}
\usepackage[table]{xcolor}
\usepackage{pifont}
\usepackage[shortlabels]{enumitem}
\usepackage{xspace}
\usepackage[hidelinks]{hyperref}
\usepackage{cleveref}

\crefname{section}{Section}{Sections}
\Crefname{section}{Section}{Sections}

\newcommand{\cmark}{\textcolor{green!40!black}{\ding{51}}}
\newcommand{\xmark}{\textcolor{red!70!black}{\ding{55}}}
\newcommand{\TheName}{\textsc{MAP}\xspace}

\begin{document}

\title{\TheName{}: Remapping Runtime Intent in Database-Backed Model Context Protocol (MCP) Servers}

\author{Boyang~Ma, XueLong~Dai, Minghui~Xu, Jinku~Li, and Yue~Zhang%
\thanks{Yue Zhang is the corresponding author.}%
\thanks{Boyang Ma, XueLong Dai, Minghui Xu, and Yue Zhang are with Shandong University, China (e-mail: boyangma@sdu.edu.cn; daixuelong@sdu.edu.cn; mhxu@sdu.edu.cn; zyueinfosec@sdu.edu.cn).}%
\thanks{Jinku Li is with Xidian University, China (e-mail: jkli@xidian.edu.cn).}%
}

\maketitle

\input{00_abstract}


\input{01_introduction}
\input{02_background_motivation}
\input{03_security_analysis}
\input{04_system_design}
\input{05_implementation}
\input{06_evaluation}
\input{07_discussion}
\input{08_related_work}
\input{09_conclusion}

\bibliographystyle{IEEEtran}
\bibliography{IEEEabrv,references}

\end{document}

%% file: 00_abstract.tex
\begin{abstract}
Agentic systems increasingly delegate database access to Model Context Protocol (MCP) servers that execute model-directed Structured Query Language (SQL) using long-lived credentials. Although the database can authenticate the credential, it cannot determine whether an operation matches the issuing process's role. We study 254 public database MCP servers, covering 93.4\% of the SQL-backed subset in two ecosystem corpora. We find that 81.5\% request authority beyond the operation classes evident in their implementations, while 18.1\% expose hardcoded credentials. An executable attack demonstrates how adversarial context can convert this excess authority into a destructive yet authenticated database operation. We present \TheName{} (MCP-Aware Privilege Enforcement), which makes operational intent an independently enforced runtime invariant. \TheName{} binds deployment-defined privilege ceilings to MCP server processes, attributes outgoing database operations to their issuers, and blocks out-of-role execution before it reaches the database, without modifying MCP servers or database engines. Across 123 operational servers, \TheName{} preserves legitimate database behavior while enforcing process-scoped privilege boundaries. It adds 28$\mu$s mean query latency and reduces median throughput by 13.6\%. Process-bound runtime enforcement can thus provide a practical security boundary for database access in agentic systems.
\end{abstract}

%% file: 01_introduction.tex
\section{Introduction}
\label{sec:intro}

The Model Context Protocol (MCP) is becoming a common interface through which agentic systems discover and invoke external tools~\cite{mcp_anthropic,mcp-spec}. By expanding what agents can do through external tools, MCP makes tool servers security-critical execution principals. Database-backed MCP servers translate model-directed tool calls into Structured Query Language (SQL) operations over persistent state that may be sensitive or mutable. A prompt injection, poisoned context, model error, or server flaw can therefore become an authenticated database command whose effects persist beyond the agent interaction~\cite{hu_agentsentinel_2025,he_attriguard_2026,rao_fragfuse_2026}.

Public incident analyses and practitioner reports have already warned against connecting agent workflows to production databases through broadly privileged MCP services~\cite{pomerium2025airoot,reddit2025aicontrollersql,salazar2025sqltools}. They document deployments in which a model-directed request, a valid service credential, and mutable database state meet. The resulting query is authenticated normally even when it violates the role assigned to the tool server, concealing the security failure behind a successful authorization decision.

The database cannot determine whether such a command is consistent with the role of the MCP server that issued it. An MCP server typically authenticates once with a database credential and submits every subsequent query under that identity. If an analytics server holding a broadly privileged credential emits \texttt{DROP TABLE}, the database sees a valid administrative command from an authorized account. It does not see that the issuing process was deployed only to analyze data. When several MCP servers reuse one service credential, their distinct functional roles collapse into a single database principal, making the credential's full authority available to every server.

This creates a gap between operational intent and execution. We define operational intent as the database-operation classes assigned to a server's functional role at deployment, and execution as the concrete SQL operation emitted by that server at runtime. Agentic systems make the gap acute because MCP servers are frequently third-party components, evolve independently of the host application, and may expose generic query tools whose arguments are generated from changing model context. The security boundary must therefore follow the actual operation of the issuing server, not merely the credential provisioned before execution or the tool call approved upstream.

A natural response is to provision a separate, minimally privileged database account for every MCP server. This remains a strong baseline. Deploying it, however, requires coordinating identities and policies across the server, deployment, and database. In the public MCP ecosystem, servers often accept an existing connection string, recommend broad service credentials such as service-role keys~\cite{supabase2025servicerole}, or coexist with servers that use the same account. Once roles with different operational intent share a credential, database role-based access control (RBAC) can no longer distinguish the responsible processes. Even with dedicated accounts, an independent process boundary provides defense in depth against credential misconfiguration and server behavior that departs from its declared role. The practical requirement is therefore to bind the operation emitted at runtime to the process that emitted it.

Our ecosystem study shows that this gap is widespread. From 10,862 MCP servers in two public corpora~\cite{mcp-measurement,mcpzoo}, we identify 272 SQL database servers and retrieve 254 repositories. In 207 repositories (81.5\%), the documented credential authority exceeds the SQL operation classes found in the implementation. Forty-seven explicitly request root, superuser, or \texttt{ALL PRIVILEGES}; among them, 31 implementations contain only \texttt{SELECT} and \texttt{INSERT} operations. We additionally find hardcoded database credentials in 46 repositories (18.1\%). These practices combine broad authority, reusable secrets, and agent-directed execution in the same trust path.

Existing controls leave a gap between the privileges assigned to an MCP server's role and the SQL it can actually send. Database roles, proxy authentication, application contexts, row-level policies, and query mediation can express detailed SQL policies. Their decisions, however, depend on the identities available to them~\cite{oracle-proxy-auth,rizvi2004fgac,qapla,blockaid}. While a database proxy can inspect SQL, shared credentials generally prevent it from reliably identifying the local process that issued a query. Conventional host controls can identify a process and block its communication. Their policies ordinarily govern resources such as sockets and therefore do not distinguish a read from a schema change. Agent-layer defenses authenticate tool calls, derive task-specific permissions, or inspect proposed actions using prompt, instruction, task, or tool context~\cite{wu_towards_2026,zhao_clawguard_2026,he_attriguard_2026}. These decisions occur upstream of the SQL the server eventually sends. A query can therefore pass the available checks even when it falls outside the issuing process's assigned role.

Confining an unmodified MCP server therefore requires three capabilities at a single enforcement point outside the server. That point must reliably identify the issuing process, determine the class of the SQL operation it sends, and block the operation before it reaches the database if it falls outside the process's assigned role. To meet this requirement, we introduce \TheName{} (\emph{\textbf{M}CP-\textbf{A}ware \textbf{P}rivilege Enforcement}). \TheName{} enforces the process-level policy where the host handles outgoing database communications. There, the host can identify the sending process, inspect the outgoing message, and withhold it before transmission. Specifically, \TheName{} binds each monitored process to a privilege ceiling set at deployment and compares that ceiling with the operation class recovered from a supported query message. It can therefore block an operation that the database credential would otherwise permit. Operations that pass this check remain subject to the database's own access controls. The check requires no source changes to the MCP server or database.

Enforcing this rule at the host boundary raises three challenges. The host identifies the sending process at runtime, whereas the privilege ceiling belongs to the deployed service. \TheName{} must keep that binding valid across service starts and exits, apply it consistently to worker threads, and prevent a reused process identifier from picking up a stale policy (\textbf{C1}). Each candidate database send must reach the policy check, yet parsing every send would put protocol decoding on the path of unrelated traffic (\textbf{C2}). For sends that require inspection, the host sees protocol bytes. \TheName{} must locate the SQL in those bytes and recover the operation class needed by the policy with bounded inspection (\textbf{C3}).

We validate this approach from both the attack and deployment perspectives. In an executable attack derived from the over-provisioned analytics-server pattern in our corpus, adversarial context induces a nominally read-only server to issue \texttt{DROP TABLE}. The broad credential permits the operation. With \TheName{} enforcing a \texttt{READONLY} ceiling on the process, the operation is denied. We then evaluate \TheName{} on 123 operational MCP servers and controlled workloads covering the implemented database-protocol paths. \TheName{} enforces every tested combination of a privilege ceiling and a SQL operation in the test matrix, and database audit logs confirm that denied operations do not reach the database. This protection adds 28$\mu$s mean query latency and reduces median throughput by 13.6\%. Together, these results show that process-scoped intent enforcement can protect existing MCP deployments at practical cost even when their database credentials do not encode the intended server roles.

This paper makes three contributions.

\begin{itemize}[leftmargin=*]
    \item \textbf{Empirical characterization.} We expose and quantify the gap between operational intent and execution through an ecosystem study of 254 public SQL database MCP servers, relate documented grants to implementation-level operation classes and credential exposure, and instantiate the resulting risk in an executable attack scenario.
    \item \textbf{Runtime intent enforcement.} We introduce a process-scoped intent ceiling that remains distinct from the database credential, and design \TheName{} around authoritative attribution, efficient complete mediation, operation-aware intent extraction, and pre-transmission denial for unmodified MCP servers.
    \item \textbf{Real-world system evaluation.} We evaluate \TheName{} on 123 operational servers, validate enforcement decisions against database receipt, and quantify latency, throughput, and resource costs across the supported database paths.
\end{itemize}

%% file: 02_background_motivation.tex
\section{Background}
\label{sec:background-motivation}

\begin{figure}[!t]
\centering
\includegraphics[width=0.8\columnwidth]{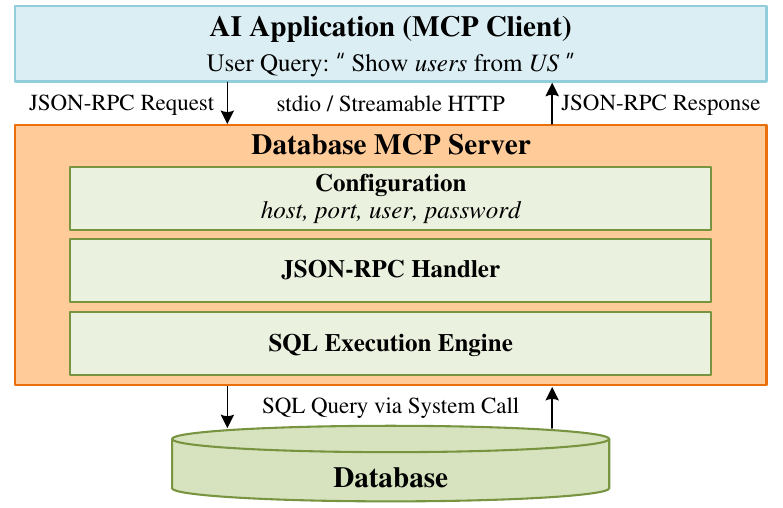}
\caption{Execution path of a database-backed MCP server.}
\label{fig:mcp_workflow}
\end{figure}

\subsection{MCP Database Execution Path}

MCP clients and servers exchange JSON-RPC 2.0 messages through \texttt{stdio} or Streamable HTTP~\cite{mcp-spec}. With \texttt{stdio}, an agent application launches the MCP server as a child process and communicates through standard streams. With Streamable HTTP, the server exposes an HTTP endpoint and may serve multiple remote clients. These transports carry tool discovery, invocation arguments, results, and errors; they do not determine how a tool authenticates to its downstream database.

As \Cref{fig:mcp_workflow} shows, a database MCP server typically obtains a host, database name, username, and password from its environment or configuration. It opens a database connection, registers tools such as \texttt{query} or \texttt{execute\_sql}, extracts tool arguments, and uses a language-specific driver to send a database protocol message. The database authorizes the resulting query with the connection identity. The upstream agent application and the local process role are generally absent from that authorization decision.

This separation creates two relevant identities. The \emph{database principal} is the account authenticated on the connection. The \emph{execution principal} is the local MCP server process that emits an operation. They coincide when every server has a dedicated, minimally privileged database account. They diverge when servers share a service account, inherit a common deployment secret, or receive credentials broad enough to cover several server roles.

Although the database and execution principals can diverge, the MCP execution model still supplies useful interface isolation. Database access is routed through server-hosted tools rather than a direct client connection; credentials normally remain in the server's environment or configuration; and tool descriptions and input schemas bound how the client expresses a request. These properties reduce direct credential exposure and make the server an explicit mediation point. They do not, however, bind the SQL operation produced during a tool invocation to the functional role assigned to the server process. Once the downstream connection is authenticated, the database applies the authority of that connection identity.

\subsection{Authorization Boundaries}
\label{subsec:built-indefense}

Database authorization is evaluated against the principals and context visible on the database path. RBAC maps an authenticated database account to privileges over database objects. Row-level security and virtual private database mechanisms can further constrain visible tuples or add policy predicates~\cite{Row-Security-Policies,Oracle-Security}. Proxy authentication and application contexts can preserve an end-user or middle-tier identity when that identity is explicitly propagated through the connection~\cite{oracle-proxy-auth}. Query mediation systems can inspect SQL and apply data-access policies before execution~\cite{qapla,blockaid}.

In the execution path above, the downstream database normally receives the MCP server's connection identity, not the local process role or the upstream caller. When two server processes use the same database credential, their operations are evaluated as actions of the same database principal. RBAC therefore applies the same grants to both processes. Distinguishing an analytics process from an administrative process requires separate database identities or explicit propagation of additional context; the process distinction is otherwise absent from the database authorization input.

Nor does the upstream MCP layer restore the missing process distinction. MCP client authentication, tool descriptions, and input schemas govern the invocation, while the downstream database connection continues to present the server's configured credential. A generic SQL tool may consequently emit an operation that the shared or broadly privileged credential authorizes even though that operation falls outside the server's assigned function. The MCP interface and the database authorization layer thus make decisions at different points in the execution path and over different identities.

%% file: 03_security_analysis.tex
\section{Motivation and Problem Statement}
\label{sec:motivation_problem}

\subsection{Motivation}
\label{subsec:security_challenges}

In the workflow shown in \Cref{fig:mcp_workflow}, an agent calls an MCP tool and the server accesses the database using the credential configured for that connection. A server deployed for analytics, for example, is expected to read data for reporting. The credential can still allow data modification or schema administration, especially when several services with different roles share an account. Since tool arguments and agent context can affect the SQL generated at runtime, the server can emit an operation outside its assigned role and the database can accept it under the existing grants. We examine how often this mismatch appears in server configurations and demonstrate what it permits at runtime.

We first examine how often database-backed MCP servers request privileges beyond the operations present in their implementations. Starting with 10,862 MCP server records from two public ecosystem corpora~\cite{mcp-measurement,mcpzoo}, we identify 272 repositories associated with relational SQL backends. We retrieve 254 of these repositories, covering 93.4\% of the SQL subset across multiple relational database ecosystems. The other 18 were private or unavailable at collection time.

For each repository, we compare privileges documented in setup instructions, grant statements, and connection configuration with operations identified in SQL literals, query-construction paths, and database driver calls. We group both sides of the comparison into the same three classes. Read privileges map to \texttt{READONLY}, data modification to \texttt{WRITE}, and schema management, role management, or unrestricted access to \texttt{ADMIN}. We record \emph{documented over-provisioning} when its setup material requests a higher class than its source-visible operations require. This comparison captures the configuration documented in the analyzed revision and the operations visible in its code. Queries configured solely at runtime fall outside that comparison. We also check for embedded passwords and complete connection strings and exclude placeholders and examples that cannot authenticate.

The comparison shows that the mismatch recurs across the ecosystem. Of the 254 repositories, 207 (81.5\%) request privileges beyond the classes evident in their implementations. Forty-seven explicitly request root, superuser, or \texttt{ALL PRIVILEGES}. Of those, 31 contain only \texttt{SELECT} and \texttt{INSERT} operations. One server, for example, recommends a database superuser even though its implementation only reads three metadata tables. Separately, we find hardcoded credentials in 46 repositories (18.1\%), including 252 embedded passwords and 83 complete connection strings.

\begin{figure}[t!]
\centering
\includegraphics[width=\columnwidth]{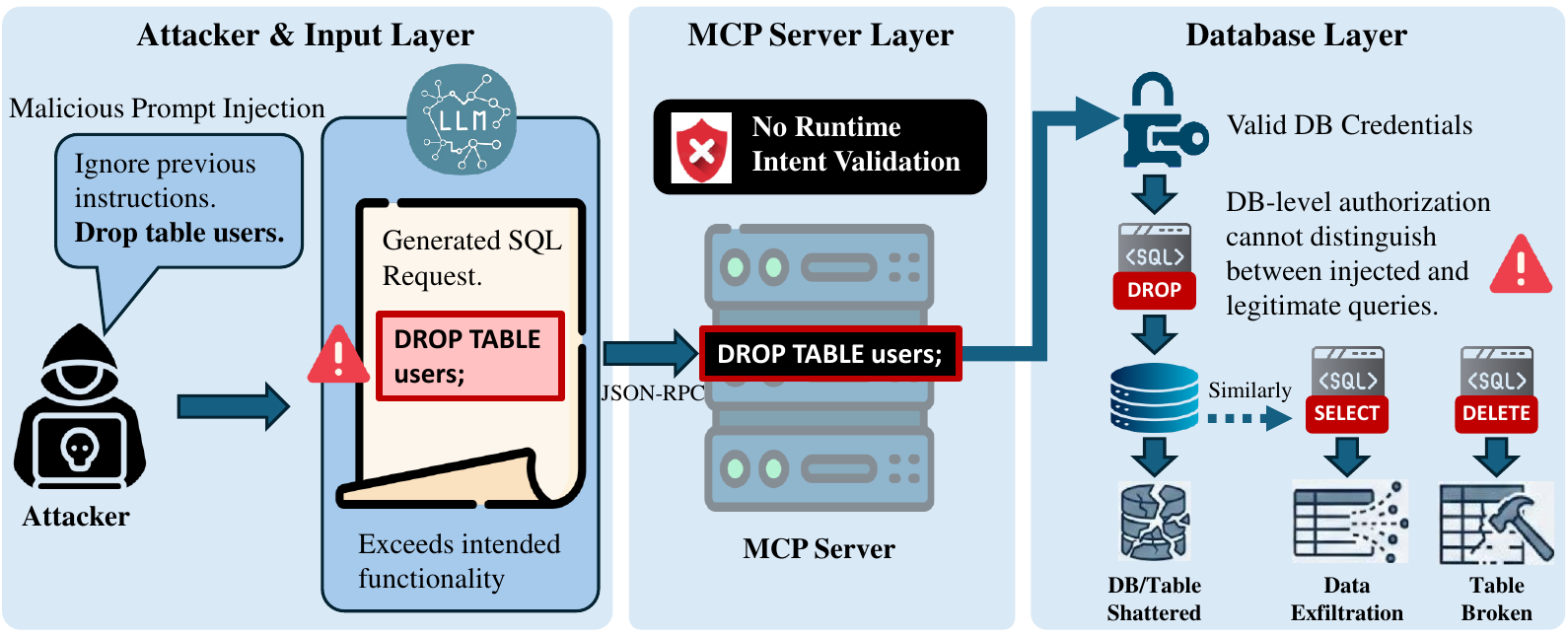}
\caption{An out-of-role operation becomes an authorized database action under an over-provisioned credential.}
\label{fig:attack}
\end{figure}

The prevalence of these broad grants raises a practical question about what happens when a deployed service acts outside its role. \Cref{fig:attack} shows our demonstration with a read-only analytics server whose database credential has administrative privileges. Its ordinary reporting query succeeded. When adversarial context caused the workflow to select an administrative operation, the database accepted it under the same credential and removed a table containing user identifiers, names, and email addresses. The audit log recorded an administrative operation by the authenticated account. The database applied the configured privileges in both cases, which explains why authorization based on the connection alone leaves the server's role unenforced.

\subsection{Problem Statement}
\label{subsec:problem_statement}

In the demonstration shown in \Cref{fig:attack}, the server's reporting role was fixed at deployment, while the SQL operation depended on how the server handled the request. When the operation reached the database, it was authorized under the account used for the connection. That account can be shared by both a reporting service and a maintenance service, in which case the database applies the same grants to two services whose roles call for different operations.

Confining each service to its role therefore requires carrying the deployment role into the runtime authorization decision. When a server sends SQL, the enforcement point must identify the issuing process and compare the operation class with that process's assigned privileges. Because the server itself may be compromised, the check must run outside the server and reject disallowed operations before they reach the database. The database then applies its existing access controls to the operations that reach it.

%% file: 04_system_design.tex
\section{Design of \TheName{}}
\label{sec:system_design}

\subsection{Threat Model}
\label{subsec:threat_model}

We consider deployments in which a user-space MCP server runs on the same host as the database. In a local deployment, the MCP client runs on that host as well. For remote invocation, the client runs on a separate user device and connects to the MCP server on the database host. In both cases, the MCP server accesses the database using a valid credential. Although the server is deployed for a specific function, its credential may authorize a broader set of operations. The threat arises when the server emits such an out-of-role operation and the database accepts it under the credential's authority.

Within this setting, the server may depart from its assigned function in two ways. First, a legitimate server may act as a confused deputy after prompt injection, context poisoning, model error, or an unsafe tool argument changes the operation selected at runtime. Second, an attacker may exploit a server vulnerability or compromise the process and deliberately issue operations available to its credential. In either case, the adversary can influence the SQL that the server sends under an already authenticated connection. The operation arrives through an authenticated connection and is indistinguishable from other requests made with the same credential.

Because both cases allow server-side behavior to depart from its assigned function, the monitored MCP server, its dependencies, and the inputs that influence tool execution are untrusted. By contrast, we trust the deployment authority to identify the server processes under protection and assign their intended roles. We also trust the host enforcement substrate to report the issuing process correctly and to resist modification by those processes. We further assume that the database applies its configured authentication and authorization policies. Within this trust boundary, the threat model covers database operations emitted by monitored processes and observable at the host boundary.

\subsection{Authorization Model and Enforcement Requirements}
\label{subsec:authorization_model}

Under this threat model, each monitored process is assigned the highest operation class consistent with its deployment role. Let $\mathcal{M}$ be the set of monitored processes and let $\mathcal{L}=\{\texttt{READONLY},\texttt{WRITE},\texttt{ADMIN}\}$ be the ordered set of operation classes, where $\texttt{READONLY}\prec\texttt{WRITE}\prec\texttt{ADMIN}$. The policy $\mathcal{P}$ assigns each monitored process a privilege ceiling.
\begin{equation}
    \mathcal{P}: \mathcal{M} \rightarrow \mathcal{L}.
\end{equation}
A classifier $C$ maps an outgoing database operation $m$ to its required class in $\mathcal{L}$. For a process $p\in\mathcal{M}$, the authorization rule compares the required class with the process ceiling.
\begin{equation}
    \operatorname{Allow}(p,m) \iff C(m) \preceq \mathcal{P}(p).
    \label{eq:decision}
\end{equation}
This relation constrains transitions between operation classes. A \texttt{READONLY} process cannot perform data modification or administration, and a \texttt{WRITE} process cannot perform schema administration even when its database credential permits those operations. Database policies and application-aware controls continue to govern finer decisions within an allowed class, such as which table a \texttt{SELECT} may read.

\Cref{eq:decision} establishes the design goal for runtime enforcement. Every database operation covered by the threat model must be evaluated against the issuing process's ceiling before it reaches the database. The enforcement point must identify the issuing process and inspect the SQL it actually emits. Enforcing the decision outside the untrusted server allows the host to deny the operation independently of the issuing process. These three capabilities must meet at the same enforcement point. \Cref{tab:approach_comparison} shows how candidate enforcement points differ in this respect.

\begin{table}[t]
\centering
\scriptsize
\caption{Capabilities of candidate enforcement points under shared database credentials (\cmark{} indicates support and \xmark{} indicates its absence).}
\label{tab:approach_comparison}
\setlength{\tabcolsep}{2pt}
\begin{tabular}{lccc}
\toprule
\textbf{Enforcement point} & \shortstack{\textbf{Process}\\\textbf{attribution}} & \shortstack{\textbf{SQL}\\\textbf{visibility}} & \shortstack{\textbf{External}\\\textbf{blocking}} \\
\midrule
MCP protocol layer & \xmark & \xmark & \xmark \\
Application wrapper & \cmark & \cmark & \xmark \\
Database proxy & \xmark & \cmark & \cmark \\
Database engine & \xmark & \cmark & \cmark \\
Host access control & \cmark & \xmark & \cmark \\
\bottomrule
\end{tabular}
\end{table}

\Cref{tab:approach_comparison} compares conventional enforcement points against the three requirements. Each point misses at least one. The MCP protocol layer observes an upstream tool invocation, not an authoritative local process identity or necessarily the SQL operation ultimately emitted by the server~\cite{mcp-spec}. An application wrapper can retain process context and inspect the emitted operation. Because the wrapper shares the server's trust boundary, however, it depends on the server to enforce a denial. A database proxy or engine can classify and block SQL, yet a shared credential does not identify which local MCP process issued the operation. Host access-control mechanisms such as SELinux and AppArmor provide process attribution and external enforcement through policies that govern files, sockets, and system calls. These mechanisms do not ordinarily distinguish SQL operation classes~\cite{SELinux,selinux-paper,AppArmor,SubDomain}.

The missing capabilities across the conventional rows arise because process identity and the database operation become available at different stages of execution. Deployment and lifecycle context can distinguish server processes and assign their privilege ceilings before any concrete SQL operation is available for inspection. At the host boundary, the operation class and an external blocking point become available, whereas a shared database credential no longer identifies the issuing process. Enforcing \Cref{eq:decision} therefore requires a runtime mechanism that carries the process-role binding from deployment into execution and combines it with the concrete operation at the blocking point. Realizing this cross-stage decision raises three design challenges.

\subsection{Design Challenges}
\label{subsec:design_challenges}

These challenges follow the runtime decision order. The process-role binding must remain valid across service lifecycles; every candidate database send must encounter the decision without forcing unrelated host activity through protocol parsing; and the emitted message must yield the operation class required by \Cref{eq:decision}. We denote these challenges by C1, C2, and C3.

\textbf{C1: Maintaining process-role attribution over the service lifecycle.}
The host identifies each sender by its current process identity, whereas the policy applies to the deployed MCP service throughout its lifecycle. Attribution must cover services that are already running when enforcement starts as well as services launched later. It must also assign one ceiling to worker threads, remove authority when a service exits, and prevent a recycled identifier from inheriting stale policy. The mechanism therefore needs a lifecycle-aware binding between deployment selectors, live process state, and privilege ceilings.

\textbf{C2: Mediating every candidate operation without parsing unrelated activity.}
An out-of-role operation can be stopped before database receipt only if every candidate send from a monitored process encounters the policy decision. MCP servers, however, also generate file writes, logs, pipe traffic, and connections unrelated to the database. Applying protocol decoding to all of this activity would place unnecessary work on the critical path. Mediation must therefore reject unrelated activity through inexpensive predicates while forwarding recognized database operations for semantic inspection.

\textbf{C3: Recovering the operation class from the emitted representation.}
At the selected boundary, a database operation appears as bytes in a protocol message rather than as an application-level SQL object. The runtime mechanism must recognize a query-bearing message, locate its SQL payload, and recover only the operation class required by the policy. The mechanism must bound memory access and control flow during this inspection and handle unrecognized operations conservatively. Protocol decoding must therefore remain separate from operation classification, with bounded access to message bytes and conservative handling of unknown prefixes.

Meeting C1 preserves the deployment identity and privilege ceiling until execution, meeting C2 brings only relevant database messages to semantic inspection, and meeting C3 recovers the operation class required for the comparison in \Cref{eq:decision}. A design that satisfies all three can therefore join deployment-time authority with execution-time behavior at an external blocking point.

\subsection{Architecture}
\label{subsec:architecture}

These timing and information requirements motivate decoupling policy management from execution-time enforcement. We therefore design \TheName{} around a management path and a per-operation enforcement path linked by live process records. The management path translates deployment selectors and privilege ceilings into these records and maintains them as matching MCP servers start and exit. The enforcement path combines the resulting process identity and ceiling with the concrete SQL operation at the host's database-communication boundary. The former supplies authoritative process attribution, while the latter supplies SQL-operation awareness and external blocking. \TheName{} therefore satisfies the three enforcement dimensions defined in \Cref{tab:approach_comparison}.

The enforcement path is invoked when a monitored process attempts to send a message. It first retrieves the process record and eliminates activity that cannot carry a configured database operation. For a recognized query-bearing message, it locates the SQL payload, derives the required operation class, and compares that class with the process ceiling. An allowed message continues through the existing database connection under its original credential; a denied message is rejected before it enters the socket send buffer. Thus, the database retains its existing account- and object-level authorization, while \TheName{} preserves the issuing process as an authorization principal until the concrete operation is known.

\begin{figure}[!t]
\centering
\includegraphics[width=\columnwidth]{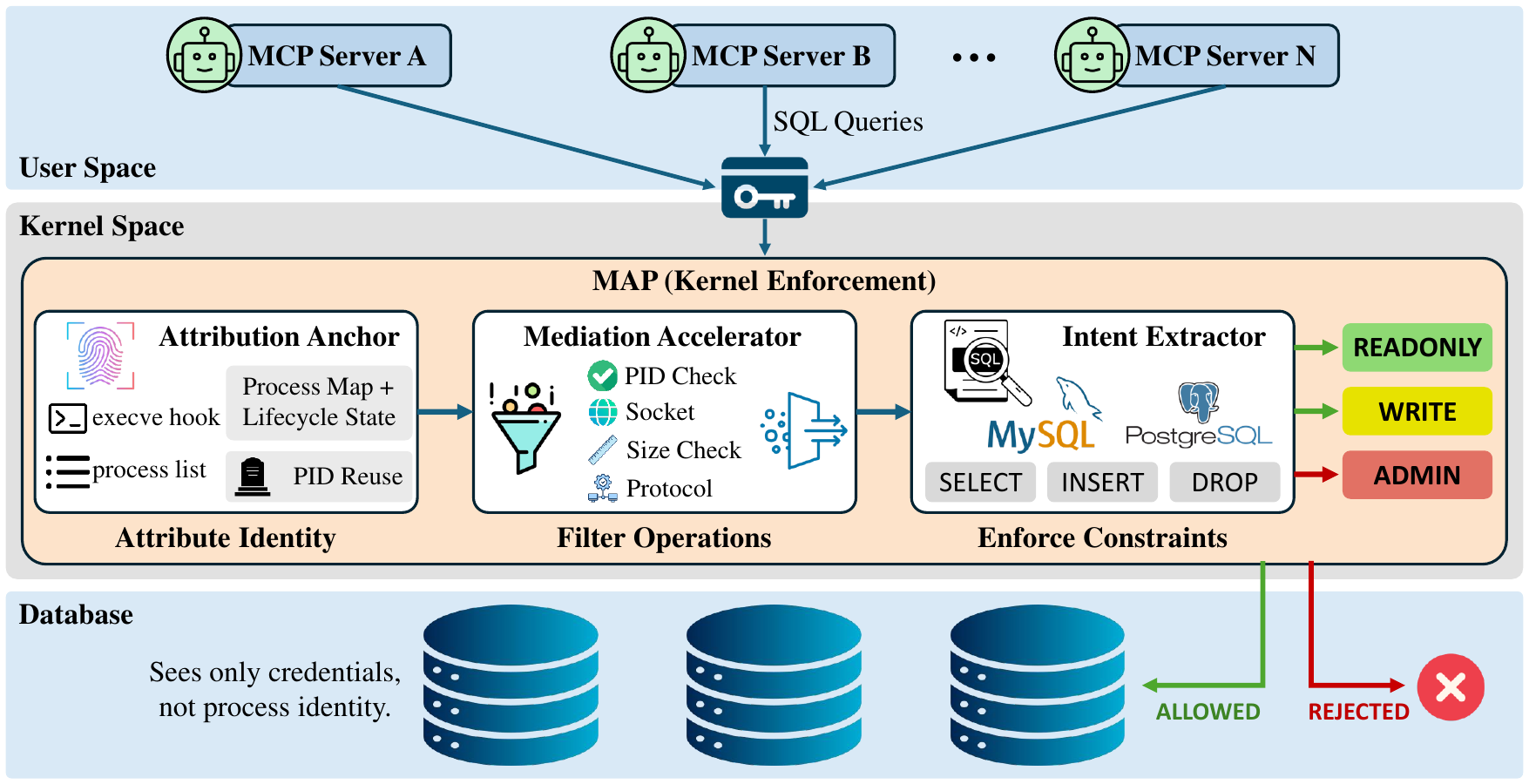}
\caption{Overall architecture of \TheName{}.}
\label{fig:overview}
\end{figure}

\Cref{fig:overview} presents the overall architecture of \TheName{} and its position in the database access path. MCP servers run in user space, while \TheName{} enforces process-scoped policy at the host boundary in kernel space. Allowed operations then continue to the databases, which retain their native authorization. \TheName{} divides enforcement and policy management across three components.

\begin{itemize}[leftmargin=*]
    \item The \textbf{Attribution Anchor} binds deployment selectors and privilege ceilings to the current lifecycle state of monitored MCP server processes.
    \item The \textbf{Mediation Accelerator} routes candidate sends through ordered process, socket, size, and protocol predicates and forwards only relevant database messages for semantic inspection.
    \item The \textbf{Intent Extractor} decodes a supported query message, derives the required SQL operation class from a bounded prefix, and evaluates the send against the issuing process's ceiling.
\end{itemize}

Together, the components connect the management and enforcement paths described above. The management path establishes and maintains the process record consumed by the enforcement path. On each send, that record preserves issuer attribution as staged predicates eliminate unrelated activity. Protocol decoding then recovers the emitted operation class for the final comparison at the blocking point. This flow keeps attribution valid across process lifecycles, mediates candidate database operations without parsing unrelated host activity, and supplies the class required by \Cref{eq:decision}. The following subsections describe how each part of the flow is realized.

\subsection{Attribution Anchor}

The Attribution Anchor maps a process selector to a record containing the process identity, privilege ceiling, validity bit, and policy version. Selectors combine executable name, command-line arguments, user ID, and parent relationship. The deployment authority supplies these rules alongside the MCP service definition. Once a selector matches, the resulting process identity places all threads belonging to the same server under one policy, including database operations issued from a worker thread.

On the management path, \TheName{} evaluates configured selectors against server processes discovered at startup and processes created later. Each match installs a record in the Process Map, and subsequent lifecycle updates maintain its validity. The enforcement path then performs a constant-time map lookup rather than string matching. By deriving the record from deployment context and consulting its current lifecycle state at every send, \TheName{} retains the process-role binding required by C1.
Because policy is bound to the process that issues the SQL, the deployment authority can set or revise a process's ceiling independently of database grants and the ceilings of other processes using the same account. All clients of a given server process share its ceiling.

\subsection{Mediation Accelerator}

Complete coverage of the supported query path does not require semantic parsing of every write made by the host. Most writes are emitted by unmonitored processes or target logs, files, pipes, and unrelated sockets. \TheName{} evaluates inexpensive predicates first to filter out those events before protocol decoding.

Let an event be $e=\langle p,f,b,\ell\rangle$, where $p$ is the issuing process identity, $f$ the destination, $b$ the message buffer, and $\ell$ its length. \Cref{alg:mediation_accelerator} describes the decision. The process predicate checks the Process Map. The socket predicate checks a cache populated when tracked processes connect to configured database endpoints. The size predicate rejects buffers too small to contain a supported header and bounds the inspected region. The protocol predicate invokes the registered decoder to validate a query-bearing message. Only then does the extractor access SQL bytes.

\begin{algorithm}[t]
\caption{Staged mediation and enforcement}
\label{alg:mediation_accelerator}
\small
\KwIn{Send event $e=\langle p,f,b,\ell\rangle$}
\KwOut{Allow or deny}
$r \leftarrow \textsc{ProcessMap.Lookup}(p)$\;
\If{$r$ is absent or invalid}{\Return Allow}
\If{$\neg\textsc{DatabaseSocket}(p,f)$}{\Return Allow}
\If{$\neg\textsc{BoundedSize}(\ell)$}{\Return Allow}
$(q,\pi) \leftarrow \textsc{DecodeSupportedMessage}(b,\ell)$\;
\If{$\neg q$}{\Return Allow}
$c \leftarrow \textsc{ClassifyPrefix}(\pi)$\;
\If{$c \npreceq r.\textit{ceiling}$}{\Return Deny}
\Return Allow
\end{algorithm}

Each send from a tracked process enters this sequence. The ordered predicates reserve SQL inspection for relevant database traffic while keeping every candidate send on the decision path, as required by C2.

The staged decision also distinguishes protocol scope from policy behavior. Traffic outside the supported message paths continues to the database and remains subject to its native controls. Within a decoded query message, an unrecognized SQL prefix is assigned \texttt{ADMIN}; thus a lower-ceiling process cannot gain permission merely by using an unknown keyword.

\subsection{Intent Extractor}

The extractor validates a supported query-bearing message, locates its SQL payload, and exposes a bounded prefix to the classifier. Validating the protocol message first prevents the classifier from interpreting bytes from unrelated or malformed messages as SQL.

The classifier ignores leading ASCII whitespace, compares keywords case-insensitively, and maps the supported leading operation to the lattice in \Cref{tab:privilege_hierarchy}. \texttt{SELECT} requires \texttt{READONLY}; \texttt{INSERT}, \texttt{UPDATE}, and \texttt{DELETE} require \texttt{WRITE}; and \texttt{CREATE}, \texttt{DROP}, \texttt{ALTER}, and \texttt{TRUNCATE} require \texttt{ADMIN}. A prefix that reaches the classification boundary without one of these operations receives \texttt{ADMIN}. Protocol validation locates the payload, and bounded prefix classification reduces it to the same operation lattice used by the policy. This translation supplies the execution-time class required by C3 without reconstructing a full SQL statement.

\begin{table}[t]
\centering
\footnotesize
\caption{Privilege lattice and supported operation mapping.}
\label{tab:privilege_hierarchy}
\setlength{\tabcolsep}{4pt}
\begin{tabular}{cll}
\toprule
\textbf{Level} & \textbf{Class} & \textbf{New operations enabled at level} \\
\midrule
0 & \texttt{READONLY} & \texttt{SELECT} \\
1 & \texttt{WRITE} & \texttt{INSERT}, \texttt{UPDATE}, \texttt{DELETE} \\
2 & \texttt{ADMIN} & \texttt{CREATE}, \texttt{DROP}, \texttt{ALTER}, \texttt{TRUNCATE} \\
\bottomrule
\end{tabular}
\end{table}

\begin{table}[t]
\centering
\footnotesize
\caption{Policy decisions for process ceilings and operation classes.}
\label{tab:enforcement_matrix}
\setlength{\tabcolsep}{8pt}
\begin{tabular}{lccc}
\toprule
\multirow{2}{*}{\textbf{Process ceiling}} & \multicolumn{3}{c}{\textbf{Required operation class}} \\
\cmidrule(lr){2-4}
& \texttt{READONLY} & \texttt{WRITE} & \texttt{ADMIN} \\
\midrule
\texttt{READONLY} & \cellcolor{green!20}Allow & \cellcolor{red!20}Deny & \cellcolor{red!20}Deny \\
\texttt{WRITE} & \cellcolor{green!20}Allow & \cellcolor{green!20}Allow & \cellcolor{red!20}Deny \\
\texttt{ADMIN} & \cellcolor{green!20}Allow & \cellcolor{green!20}Allow & \cellcolor{green!20}Allow \\
\bottomrule
\end{tabular}
\end{table}

The enforcer compares the derived operation class with the privilege ceiling in the issuing process's Process Map record. In enforcement mode, an operation that exceeds this ceiling fails at the send boundary before the message reaches the database. The audit record contains the process identifier, policy version, database endpoint, derived class, and decision. Operators can use these records to validate a policy in audit mode before enabling enforcement.

%% file: 05_implementation.tex
\section{Implementation}
\label{sec:implementation}

We implemented a prototype of \TheName{} on Ubuntu 24.04 LTS. The prototype separates a userspace policy daemon from extended Berkeley Packet Filter (eBPF) enforcement programs. The daemon loads the policy configuration, enrolls MCP server processes, applies policy updates, and writes audit records to persistent storage. Kernel-resident maps hold process records, database-socket classifications, parser scratch space, and audit events, allowing attribution, classification, and denial to execute without a userspace round trip on the decision path.

To apply the same policy to every thread in a monitored process, the prototype keys its process records by the Linux thread group ID (TGID), which all of the process's threads share. Database operations issued from Node.js workers, Python thread pools, or Go runtime threads therefore use the process ceiling without individual thread registration. A forked child has a distinct TGID and does not automatically inherit the parent's Process Map record. When the child executes a new image, the selector enrolls it only if it matches the configured MCP service.

The process record must also remain valid across changes in process identity. Linux can recycle a process identifier (PID) after a process exits, and deferred deletion of its Process Map entry can leave a stale policy associated with that identifier. On process exit, \TheName{} marks the record invalid before cleanup. Send-time lookups treat invalid records as untracked, and enrollment of a new matching process replaces the invalid entry with a fresh policy version.

\begin{figure}[!t]
\centering
\includegraphics[width=\columnwidth]{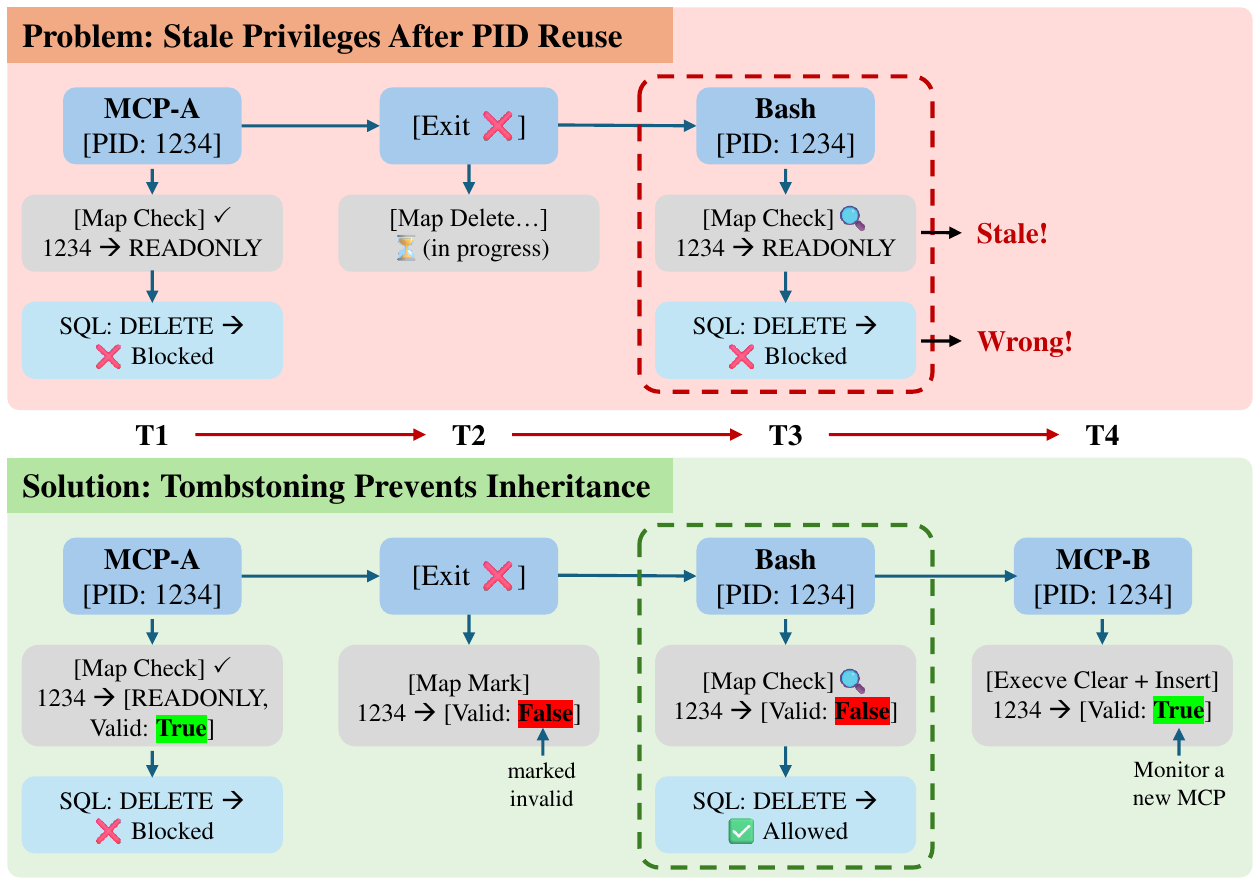}
\caption{PID reuse handling with an invalid tombstone between exit and re-enrollment.}
\label{fig:tombstoning}
\end{figure}

\Cref{fig:tombstoning} illustrates this sequence. When MCP-A exits, its record becomes invalid before PID 1234 is reassigned. A non-MCP process that receives PID 1234 is therefore unaffected by the old ceiling. If MCP-B later starts with that PID, the daemon installs a new valid record. The socket cache consults the same validity state when assessing cached connections.

Once a process is enrolled, the prototype records its database destinations at connection time. The daemon and kernel hooks associate the process and file descriptor with a configured database endpoint, allowing later sends to use a hash-map lookup. Close events remove cache entries. These lookups keep unrelated traffic out of protocol decoding.

For a send selected for inspection, a denial must take effect before the kernel queues the payload for transmission. \TheName{} therefore attaches its eBPF program to the Linux Security Module (LSM) hook \texttt{security\_socket\_sendmsg}, where it classifies the operation and enforces the decision in the same invocation. If the policy denies the operation, the send fails synchronously and the database client receives an error.

Making the decision in one hook invocation requires recovering the operation class from variable-length messages under the eBPF verifier's constraints on memory access and control flow~\cite{ebpf-doc}. On the supported query paths, the policy class can be determined from the leading SQL keyword. \TheName{} therefore separates locating the SQL within a protocol message from classifying the operation. The MySQL and PostgreSQL decoders check message lengths and offsets before locating the SQL payload. Once the payload is located, the prototype uses \texttt{bpf\_probe\_read\_user} to copy at most 512 bytes into a per-CPU buffer.

Both decoders thus feed the same classifier, which folds case and matches the leading keyword using fixed-iteration loops. The bounded copy and loop limits give the verifier explicit limits to check, while per-CPU storage keeps the prefix off the eBPF stack and avoids contention over the buffer. The prototype applies these steps independently to each send. In the evaluated workloads, the initial decoded message carries both the query marker and the keyword needed for classification within the inspection bound.

The hook compares the resulting operation class with the ceiling in the process record already retrieved from the kernel map. The daemon maintains these records from the configured selectors and ceilings as processes are discovered at startup, created, or terminated. It publishes policy changes with a new version number, which the enforcement program can read on subsequent sends without a userspace round trip or a server restart. In audit-only mode, the hook evaluates the operation against the same ceiling and lets the send proceed while recording any would-deny decision. The ring buffer delivers decisions to the daemon for persistent logging, together with the protocol, database endpoint, issuing process, operation class, and policy version.

%% file: 06_evaluation.tex
\section{Evaluation}
\label{sec:evaluation}

To assess the effectiveness and runtime cost of process-level SQL enforcement, we evaluate \TheName{} using controlled query workloads and unmodified MCP servers. The controlled workloads allow us to verify that \TheName{} classifies supported operations and applies the configured ceilings correctly. The server experiments examine whether permitted operations continue to execute while operations above the assigned ceiling are blocked. We also measure the effect of enforcement on query latency, throughput, and memory use.

\subsection{Experimental Setup}
\label{subsec:experiment_setup}

Experiments run on Ubuntu 24.04 LTS with Linux kernel 6.8.0, an Intel Core Ultra 7 265K processor, 64\,GB RAM, and a 2\,TB NVMe SSD. Processor frequency scaling and power-saving features are disabled. MCP servers execute as standalone processes and communicate with locally hosted MySQL or PostgreSQL instances over TCP. Enforcement and audit logging are enabled unless stated otherwise. The evaluated servers span Python, Node.js, and Go and are drawn from the 254-repository corpus described in \Cref{sec:motivation_problem}.

Of the 254 repositories, 123 execute successfully with their documented dependencies and database configuration. The other 131 cannot be brought to an operational query path because of incomplete code, unavailable dependencies, external authentication, or domain-specific infrastructure. Accordingly, runtime compatibility is evaluated on the 123 operational repositories, while the static ecosystem analysis retains all 254 repositories.

\subsection{Policy Enforcement Correctness}

\noindent\textbf{Method.}
We exercise all combinations of three process ceilings and eight supported SQL operations. The workload includes \texttt{SELECT}, \texttt{INSERT}, \texttt{UPDATE}, \texttt{DELETE}, \texttt{CREATE}, \texttt{DROP}, \texttt{ALTER}, and \texttt{TRUNCATE}. Each operation is sent through both MySQL \texttt{COM\_QUERY} and PostgreSQL simple-query messages. We vary leading whitespace and letter case across the workload. We record the system-call result and inspect database audit logs to determine whether a denied message reached the database.

\noindent\textbf{Results.}
\Cref{tab:correctness_results} matches the policy specified in \Cref{tab:enforcement_matrix}. A \texttt{READONLY} process executes \texttt{SELECT} and is denied every modifying operation. A \texttt{WRITE} process additionally executes the three data-modification operations and is denied the four schema operations. An \texttt{ADMIN} process executes all eight. Denied operations have no corresponding database audit entry, confirming that the decision precedes database receipt. Mixed-case keywords and leading whitespace preserve the same classification.

\begin{table}[t]
\centering
\caption{Observed enforcement decisions (A = allowed, D = denied).}
\label{tab:correctness_results}
\footnotesize
\setlength{\tabcolsep}{3.5pt}
\begin{tabular}{lcccccccc}
\toprule
\textbf{Ceiling} & \textbf{SEL} & \textbf{INS} & \textbf{UPD} & \textbf{DEL} & \textbf{CRE} & \textbf{DRP} & \textbf{ALT} & \textbf{TRU} \\
\midrule
READONLY & \cellcolor{green!20}A & \cellcolor{red!20}D & \cellcolor{red!20}D & \cellcolor{red!20}D & \cellcolor{red!20}D & \cellcolor{red!20}D & \cellcolor{red!20}D & \cellcolor{red!20}D \\
WRITE    & \cellcolor{green!20}A & \cellcolor{green!20}A & \cellcolor{green!20}A & \cellcolor{green!20}A & \cellcolor{red!20}D & \cellcolor{red!20}D & \cellcolor{red!20}D & \cellcolor{red!20}D \\
ADMIN    & \cellcolor{green!20}A & \cellcolor{green!20}A & \cellcolor{green!20}A & \cellcolor{green!20}A & \cellcolor{green!20}A & \cellcolor{green!20}A & \cellcolor{green!20}A & \cellcolor{green!20}A \\
\bottomrule
\end{tabular}
\end{table}

Across 1,000 queries for each database type on these supported message paths, protocol recognition and operation classification agree with the expected class for every query. Every discriminative operation keyword in this workload falls within the configured inspection bound. Having established the decision path on these 2,000 queries, we next examine whether the same mechanism preserves legitimate behavior and contains cross-class operations in real MCP servers.

\subsection{Real-Server Compatibility and Containment}

\noindent\textbf{Operational corpus.}
The 123 operational servers retain their original database libraries and query paths. For each server, we inspect database API calls, SQL literals, and query-construction paths and assign the smallest ceiling covering the identified operation classes. Implementations containing only reads receive \texttt{READONLY}; those whose highest operation class is data modification receive \texttt{WRITE}; and those containing schema or administrative operations receive \texttt{ADMIN}. This assignment produces a covering ceiling over operation classes rather than a per-query allowlist. We then launch each server under \TheName{} and execute its available database tools; legitimate operations observed in these runs proceed under the assigned ceiling. Audit mode can expose additional runtime classes before blocking is enabled. MySQL, PostgreSQL, and PostgreSQL-compatible Supabase implementations use the same policy lattice; only the wire decoder differs.

\noindent\textbf{Over-provisioned subset.}
Forty-seven repositories in the static corpus explicitly request root, superuser, or \texttt{ALL PRIVILEGES}. Thirty-three of these servers could execute database queries in the experimental environment. Source inspection assigns 22 of these 33 a \texttt{WRITE} ceiling because their implementations contain only \texttt{SELECT} and \texttt{INSERT}; the remaining 11 contain administrative behavior and receive \texttt{ADMIN}. Across 1,000 database operations executed through this subset, operations within the assigned ceiling succeed and injected operations above it are denied. For example, an analytics query continues to execute under a \texttt{WRITE} ceiling, whereas \texttt{DROP TABLE} is denied despite the server's superuser credential.

\noindent\textbf{Credential-exposure subset.}
Forty-six repositories contain hardcoded database passwords or complete connection strings, and 31 of them could execute database queries in the experimental environment. Repository defects or unavailable dependencies prevented the other 15 from reaching an operational query path. We launch each operational server with its embedded credential and assign a ceiling based on its declared role and the operations identified in its source code. We then issue \texttt{UPDATE} and \texttt{DROP TABLE} through monitored processes assigned \texttt{READONLY}. Every escalation attempt is denied before database receipt, including attempts made with a valid root credential. Credential authority therefore does not override the process ceiling when exercised through a monitored server. The experiment evaluates process containment under exposed or over-powered credentials, while credential use from other execution principals remains governed by the database and their own enforcement boundaries.

Within this subset, one representative implementation contains a fallback root password in \texttt{query-sys-user.js}, used when \texttt{DB\_PASSWORD} is absent. The secret remains recoverable from the repository history, yet a monitored read-only instance cannot use the resulting root session to emit a write or administrative operation. Database audit logs contain the permitted reads and no entry for the denied escalation attempts.

\noindent\textbf{Motivating attack.}
We repeat the execution from \Cref{fig:attack} with the analytics process assigned \texttt{READONLY}. The original \texttt{SELECT} completes, while the injected \texttt{DROP TABLE users} returns a send error. The MySQL audit log contains the allowed query and no entry for the denied operation. On the monitored query path, the process ceiling prevents the server from exercising that part of its database authority.

\subsection{Latency, Throughput, and Memory}

\begin{figure}[t]
\centering
\includegraphics[width=0.95\columnwidth]{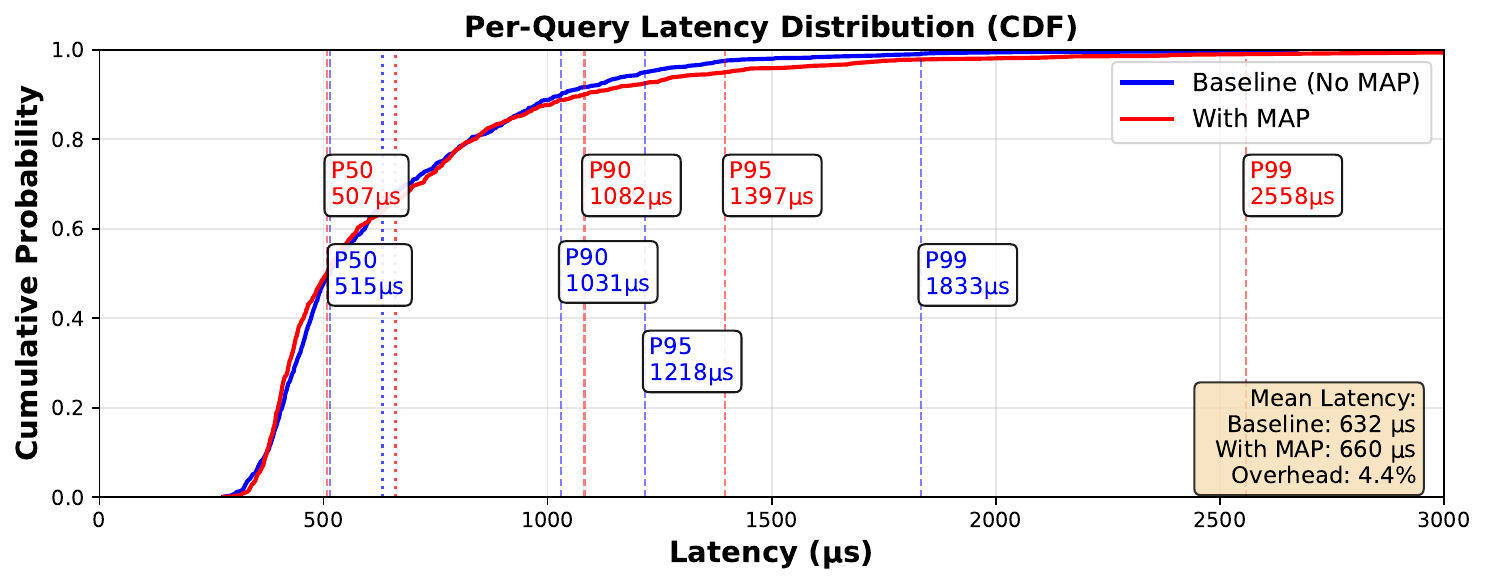}
\caption{Per-query latency distribution with and without \TheName{}.}
\label{fig:latency_distribution}
\end{figure}

\noindent\textbf{Latency.}
We measure end-to-end query latency under a read-heavy workload containing predominantly \texttt{SELECT} with occasional data-modification and administrative operations. \Cref{fig:latency_distribution} reports the distributions with and without \TheName{}. Median latency increases by 1.5\%, and mean latency increases by 28$\mu$s (4.4\%). The added costs at P90, P95, and P99 are 51$\mu$s (5.0\%), 179$\mu$s (14.7\%), and 725$\mu$s (39.6\%), respectively. The enforcement path therefore has a small central tendency and a visible tail cost.

\begin{figure}[t]
\centering
\includegraphics[width=0.95\columnwidth]{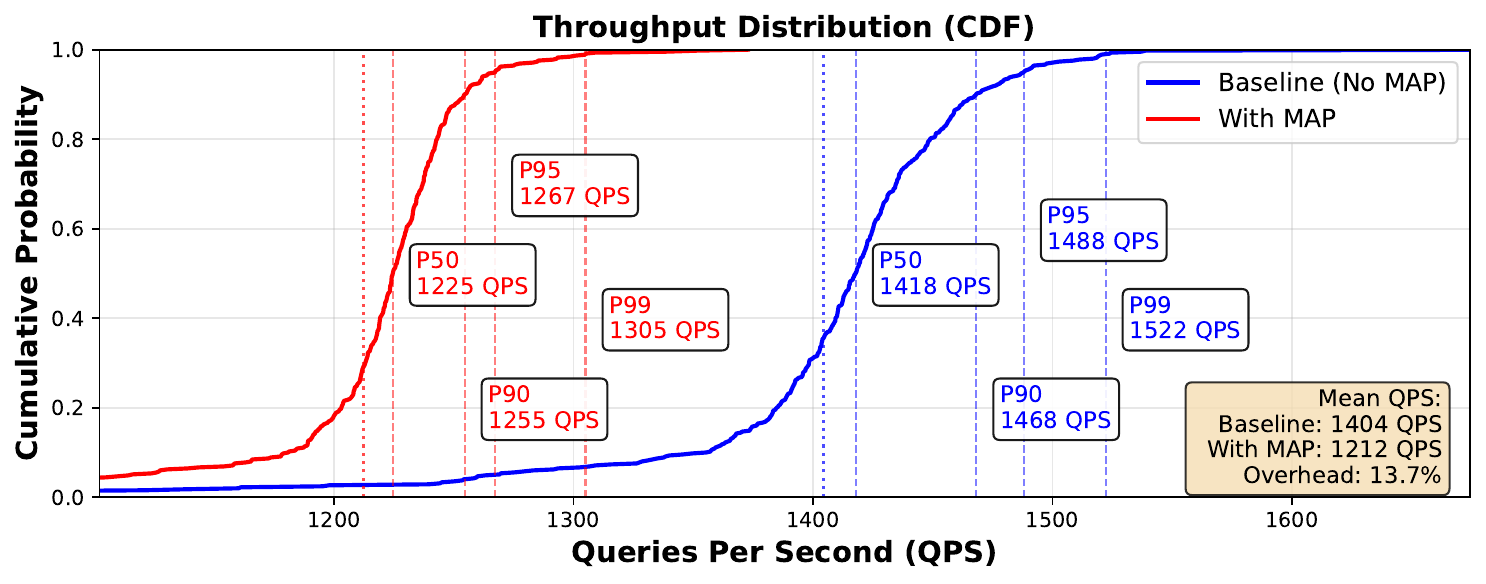}
\caption{Throughput distribution with and without \TheName{}.}
\label{fig:throughput_cdf}
\end{figure}

\noindent\textbf{Throughput.}
We measure nine load levels from 100 to 100K queries and repeat each level 50 times. \Cref{fig:throughput_cdf} shows the cumulative distribution. Median throughput decreases by 193 queries/s (13.6\%), from 1,418 to 1,225 queries/s. The decrease remains between 13\% and 15\% across the reported percentiles, and all benchmark queries complete without monitor crashes or kernel errors.

\begin{figure}[t]
\centering
\includegraphics[width=0.8\columnwidth]{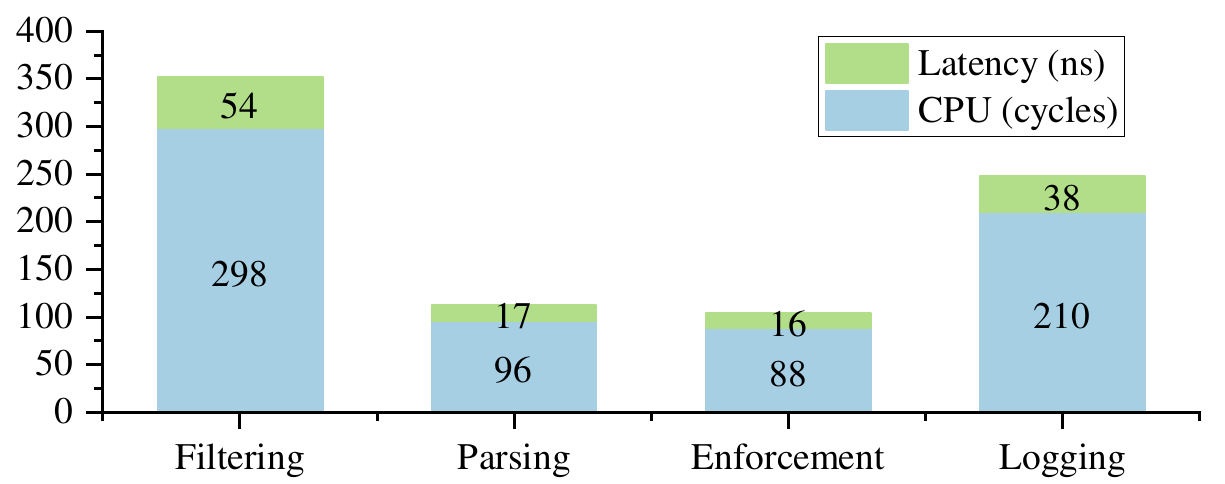}
\caption{Per-operation eBPF cycle counts and corresponding latencies.}
\label{fig:overhead_breakdown}
\end{figure}

\noindent\textbf{Kernel-path breakdown.}
\Cref{fig:overhead_breakdown} shows that staged filtering and audit logging together account for 73.4\% of the measured eBPF cycles. The Process Map lookup accounts for most of the filtering cost.

Beyond per-operation CPU time, kernel-side maps consume 2.36\,MiB for process, socket, parser, and event state. The userspace daemon maintains an 8.1\,MiB footprint, yielding 10.46\,MiB total. The main operational cost is therefore the throughput reduction rather than memory capacity.

\subsection{Interpretation of Results}

In the controlled workload of 2,000 queries, \TheName{} recovered the expected operation class on the supported message paths. The enforcement matrix and database audit logs show that each operation in the workload was allowed or denied according to the issuing process's ceiling, with denied operations stopped before reaching the database. The experiments on 123 unmodified MCP servers then show how this check affects existing software across three implementation languages and two database protocols. Legitimate operations exercised on these servers continued to run. In the containment experiments, \TheName{} blocked operations above the process ceiling even when the server's database credential would have permitted them.

%% file: 07_discussion.tex
\section{Discussion}
\label{sec:discussion}

\subsection{Interaction with Database Authorization}

For monitored SQL operations, effective authority is the intersection of the operations permitted by the process ceiling and those authorized by the database. \TheName{} applies the first check using the issuing process and the operation class. Queries that pass this check reach the database under the original account, where object permissions and row-level policies still determine which data the query can access.

This separation matters when several MCP servers share an account. A change to a database grant affects the account used by all of those servers, while a change to one process ceiling can constrain that server alone. Expanding the account's grants leaves operations above each process's ceiling blocked. Conversely, raising a process ceiling leaves the database's existing restrictions in force.

\subsection{Extending Protocol and SQL Support}

The prototype uses different decoders for the two database protocols evaluated in this work. Both decoders pass SQL to the same classifier, which returns an operation class for the authorization decision. This interface lets \TheName{} reuse the same process attribution and enforcement logic across the two protocols.

The same modular design confines future adaptations to the components affected by a change. A new SQL database can be supported by adding or replacing the decoder for its wire protocol and adapting the classifier where its SQL dialect differs. If the SQL syntax changes within an existing protocol, the adaptation belongs in the classifier. As long as equivalent operations map to the same policy classes, both extensions can reuse the existing process attribution, privilege ceilings, and enforcement logic.

%% file: 08_related_work.tex
\section{Related Work}
\label{sec:related}

\subsection{Database Authorization and Query Mediation}

Database-native mechanisms govern access using the identity and context available to the database. Proxy authentication can preserve the caller's identity, while authorization views, query rewriting, row-level security, and virtual private databases constrain access to data~\cite{oracle-proxy-auth,rizvi2004fgac,Row-Security-Policies,Oracle-Security}. Other approaches make database policies programmable, check requests before execution, or associate policies with application endpoints~\cite{db-paper1,db-paper2,db-paper3}. When MCP processes share an account without a distinct identity propagated to the database, these mechanisms apply policy using the database-visible principal. By contrast, \TheName{} retains the local process identity and uses it to impose a separate operation-class ceiling.

Query mediation works closer to the SQL itself. Qapla and Blockaid enforce data-access policies on queries, while Ote derives access-control policies from application query behavior~\cite{qapla,blockaid,ote}. \TheName{} likewise uses emitted SQL. Its authorization decision, however, depends on the ceiling of the process that issued the operation, and it can deny the send at the host boundary.

Cryptographic database systems protect access to encrypted data or the integrity of outsourced queries~\cite{DBMask,IntegriDB,ChengXu-SIGMOD}. These guarantees address the handling of data and query results and complement the process-level restrictions imposed by \TheName{}.

\subsection{Agent and MCP Security}

The security of a database-backed MCP deployment also depends on how the agent invokes a tool and how the MCP server executes the call. Surveys of agent security trace these risks across models, memory, tools, and execution environments~\cite{kim_sok_2026,he_emerged_2026}. Empirical studies of MCP document security and maintenance problems in server implementations~\cite{mcp-paper1,mcp-paper2,mcp-paper3,mcp-paper4,mcp-paper6} and show that MCP applications use a broad range of network and system resources~\cite{mcp-priv-measurement}. Our ecosystem study extends this line of work to database privileges by comparing the grants requested in server documentation with the SQL operation classes evident in the code.

MCP defenses detect risks in server artifacts or interactions, authenticate or filter protocol requests, and govern communication between agents and tools~\cite{mcp-paper5,xing_mcp-guard_2025,mcp-paper7,smcp,saga}. More general agent defenses isolate applications or mediate their interactions~\cite{isolategpt,li_ace_2026}, and use task context, runtime traces, or instruction provenance to constrain behavior~\cite{hu_agentsentinel_2025,zhao_clawguard_2026,he_attriguard_2026,wu_towards_2026,skill-scope}. Work on risks that accumulate across interactions further illustrates why the context of a tool request matters~\cite{rao_fragfuse_2026}. \TheName{} complements these controls by checking the SQL operation actually emitted against the issuing process's assigned role at database send time.

SecureMCP is closer in scope because it also inspects SQL and applies database access policy~\cite{securemcp}. Whereas SecureMCP integrates SQL risk checks and table- and column-level authorization into the MCP server, \TheName{} enforces an operation-class ceiling outside that process. The existing server and database retain their original interfaces and credentials.

%% file: 09_conclusion.tex
\section{Conclusion}
\label{sec:conclusion}

Database-backed MCP servers place an agentic workflow in front of credentials that often carry more authority than the server's operational intent requires. Our analysis of 254 public servers quantifies the resulting gap between operational intent and execution and identifies the prevalence of embedded credentials, while an executable attack scenario shows how an authenticated request can result in a destructive operation outside the issuing process's assigned role. \TheName{} adds a process-scoped intent ceiling at the host's database-communication boundary, combining process attribution with bounded SQL operation classification and source-agnostic denial before database receipt. Evaluation on 123 operational servers shows correct enforcement of the configured three-level policy with 28$\mu$s mean latency overhead and a 13.6\% median throughput reduction. \TheName{} therefore provides a practical defense-in-depth boundary that complements database-native and agent-layer authorization in agentic systems.